# Probing the topological surface states in superconducting Sn$_4$Au single crystal: A magneto transport study

M. M. Sharma[1,2], Poonam Rani[3] and V.P.S. Awana[1,2]

[1]*Academy of Scientific & Innovative Research (AcSIR), Ghaziabad, 201002, India*
[2]*CSIR- National Physical Laboratory, New Delhi, 110012, India*
[3] *Materials Science Division, Inter-University Accelerator Centre, New Delhi-110067, India*

**Abstract:**

Materials exhibiting bulk superconductivity along with magnetoresistance (MR) in their normal state have emerged as suitable candidates for topological superconductivity. In this article, we report a flux free method to synthesize single crystal of topological superconductor candidate Sn$_4$Au. The phase purity and single crystalline nature are confirmed through various characterizations viz. X-Ray diffraction (XRD), field emission scanning electron microscopy (FESEM), selected Area electron diffraction (SAED), and transmission electron microscopy (TEM). Chemical states of the constituent element viz. Sn and Au are analyzed through X-Ray photoelectron spectroscopy (XPS). Superconductivity in synthesized Sn$_4$Au single crystal is evident form ρ-T plot and critical field (H$_c$) is determined through ρ-H plot at 2K i.e., just below critical temperature T$_c$. A positive magnetoresistance (MR) is observed in ρ-H measurements at different temperatures above T$_c$, viz. at 3K, 5K, 10K and 20K. Further, the magnetoconductivity (MC) is analyzed by using Hikami-Larkin-Nagaoka (HLN) formalism, which signifies the presence of weak antilocalization (WAL) effect in Sn$_4$Au. Angle dependent magneto-transport measurement has been performed to detect the origin of observed WAL effect in Sn$_4$Au single crystal. Normalized MC vs Hcosθ plot shows presence of topological surface states (TSS) in the studied system. It is evident that Sn$_4$Au is a 2.6K topological superconductor.

*Corresponding Author
Dr. V. P. S. Awana:  E-mail: awana@nplindia.org
Ph. +91-11-45609357, Fax-+91-11-45609310
Homepage: awanavps.webs.com

**Introduction:**

The quest on search for topological materials with superconducting properties has revolutionised the field of quantum condensed matter. The unique combination of bulk superconducting gap and topological surface states enables topological superconductors to host Majorana fermions [1-4]. Recently, some of the Dirac semimetals (DSM) have gathered significant interest in this regards, as they are found to show simultaneous existence of bulk superconducting gap and topological surface sates [5-7]. DSMs have emerged as transitional state of topological trivial insulators to topological insulators (TIs) and vice versa. [8,9]. In DSMs, the Dirac point is not gapped due to intrinsic spin orbit coupling (SOC), instead this is the only point where the conduction band and valence band meet each other. These electronic bands are linearly dispersed in all directions around the Dirac point, thus forming a Dirac cone [9]. The spins of the surface electrons circulate around the Dirac cone, and create $\pi$ Berry phase in electron wavefunction [10,11]. Presence of $\pi$ Berry phase in a topological material, is considered to be the vital property of the same, which provides immunity to topological surface states against non-magnetic impurities [12]. Observation of quantum oscillations viz. Shubnikov-de Haas (SdH) or de Haas-van Alphen (dHvA) along with weak anti-localization (WAL) effect in magneto-transport measurements signify the presence of $\pi$ Berry phase in a topological material [13-16]. The electrical conduction through surface states of a TI is vulnerable to applied magnetic field, and as a consequence, the same results in occurrence of linear MR (LMR) in magneto-transport measurements [5,17,18].

Study of normal state transport properties of possible topological superconductors is an essential part to probe topological surface states. The transport phenomenon in normal state of a topological superconductor must principally be dominated by topological surface states, resulting in occurrence of significant MR above $T_c$. Some of the recently studied superconducting topological semimetals (TSMs) such as $Au_2Pb$ [5], $PdTe_2$ [17], $PbTaSe_2$ [19], $Sn_4Au$ [20] etc. have shown a large MR in their normal state, creating possibility of realization of topological superconductivity. Among these, $Sn_4Au$ is isostructural to recently discovered TSMs viz. $Sn_4Pd$ [21] and $Sn_4Pt$ [22]. Superconducting properties of $Sn_4Au$ were though discovered a long time ago [23,24], but any detailed analysis of both its superconducting and normal state properties is reported only very recently [20]. $Sn_4Au$ exhibits large MR similar to its isostructural compounds $Sn_4Pd$ and $Sn_4Pt$, which are considered to have Dirac nodal arc semimetal like band structure [21,22]. $Sn_4Au$ is also considered to lie in category of TSMs based on available theoretical reports [25,26], while experimental confirmation of the same is yet to be done. Also, $Sn_4Au$ has interesting superconducting and normal state properties as it shows 2-dimensional superconductivity with normal state MR [20]. Moreover, low temperature magneto-conductivity of Sn$_4$Au in its normal state, is contributed by topological surface states as shown in ref. 20. These exotic properties of Sn$_4$Au inspire us to study more about its superconducting and normal state properties and to compare with only available report [20], as well to further extend the analysis.

The single crystal of Sn$_4$Au is synthesized using a flux free method and is well characterized using XRD, FESEM, HRTEM and XPS measurements. ρ-T measurements shows the presence of superconductivity with onset critical temperature $T_c^{onset}$ at 2.6K. Magneto-transport measurements shows significant MR, in an applied field of ±12T. Though, in previous study on normal state properties of Sn$_4$Au, MC was fitted with simple WAL transport formula, here we used HLN equation to fit the same, and confirmed that the normal state transport is clearly dominated by topological surface states. This dominance of topological surface states is also evidenced through angle dependent magneto-transport measurements. This report clearly shows that normal state transport in Sn$_4$Au is dominated by topological surface states, while the same material shows superconductivity below its $T_c$, establishing Sn$_4$Au to be topological superconductor.

**Experimental:**

Single crystal of Sn$_4$Au is synthesized using a flux free method, here it is worth mentioning that all previous methods to synthesize Sn$_4$Au single crystal were based on flux method, in which excess of Sn amount is taken as flux. Here, high quality powders of Sn and Au were taken in stoichiometric ratio (4:1). These powders were grinded thoroughly in inert atmosphere and palletized using hydraulic press. This palletized sample was vacuum encapsulated in a quartz ampoule. This ampoule was heated to 900°C at a rate of 120°C/h. and dwelled at this elevated temperature for 24 hours. This temperature is well above the melting temperature of Sn$_4$Au, the sample is heated to a higher temperature to melt the constituent element. The melted sample is cooled to 310°C at a cooling rate of 10°C/h, and kept at 310° for 6h. Further, the sample is cooled down to 240°C at a relatively slow cooling rate of 1°C/h. This slow cooling rate provides ample time for atoms to attain their energetically stable positions, which is a necessary step in crystal growth. A hold period of 48 hours is applied at 240°C, at which stabilization of phase took place. After that the sample is allowed to cool normally to room temperature. Though, all heating protocols are same as in ref.20, the important difference is that the initial heating temperature in our case is 900°C, instead of 630°C in ref.20. This higher heating temperature removes the need of adding extra Sn as a flux. Thus, obtained crystal is silvery shiny and cleavable along its growth axis. The schematic of heat treatment is shown in Fig. 1. This method is relatively easier method to synthesize single crystal of $Sn_4Au$, than the previously suggested flux method [20], as the same does not require centrifuging of extra Sn flux.

XRD pattern on mechanically cleaved flake and powder XRD pattern were recorded using Rigaku mini flex-II table top XRD equipped with Cu-$K_\alpha$ radiation. MIRA II LMH from TESCAN made FESEM equipped with Energy-dispersive X-ray spectroscopy (EDS) is used to visualize the surface morphology and to determine elemental composition. SAED pattern and atomic planes are visualized using JEOL/JEM-F200 with One view CMOS camera (4Kx4K) Transmission Electron Microscope (TEM). XPS spectra was recorded by using XPS (Model: K-Alpha-KAN9954133, Thermo Scientific) with mono-chromated and micro-focused Al Kα radiation (1486.6eV). The spectrometer has been designed with a dual-beam flood source to provide a charge compensation option and calibration is confirmed by the position of C 1 s line at the binding energy 284.8eV. The working pressure was maintained $< 6 \times 10^{-8}$ mbar. The scans for Au 4f and Sn 3d were recorded with the step size of 0.1eV. Low temperature magneto-transport measurements were carried out using Quantum Design physical property measurement system (QD-PPMS) equipped with sample rotator. Standard four probe method is used for electrical transport measurements. Hall measurements are carried out using Vander Paw geometry. Rietveld refinement of powder XRD (PXRD) pattern is performed using Full Proof software and unit cell is drawn in VESTA software, which uses CIF obtained from Rietveld refinement.

**Result & Discussion:**

Phase purity of the synthesized $Sn_4Au$ single crystal is confirmed through Rietveld refined PXRD pattern, and the observed and fitted plots are shown in Fig. 2. Rietveld refinement is performed taking parameters of an orthorhombic crystal structure with Aba2 (41) space group symmetry. All observed peaks in PXRD pattern are well fitted with the applied parameters. Quality of fit is given by $\chi^2$ value, which is found to be 1.81 and is in an acceptable range. The lattice parameters obtained from Rietveld refinement are a=6.514(6)Å, b=6.529(1)Å & c=11.726(6)Å and α=β=γ=90°. These lattice parameters along with atomic co-ordinates are listed in table 1, along with the refinement parameters. No impurity peak is visible in PXRD pattern, this shows that the synthesized $Sn_4Au$ single crystal is phase pure. PXRD pattern is an important tool to determine the phase purity of the sample and also the same clarifies the phase in which the sample is grown. $Sn_4Au$ is reported to have two space groups viz. Ccce (68) [27] and Aba 2 (41) [20]. In ref. 27 space group of $Sn_4Au$ is determined theoretically to be Ccce (68), while in ref. 20, space group of $Sn_4Au$ is suggested to be Aba 2 (41) on the basis of XRD pattern taken on crystal flake. PXRD pattern is more effective to determine space group symmetry and phase purity. Here PXRD pattern of $Sn_4Au$, eventually confirms that $Sn_4Au$ crystallizes in orthorhombic phase with Aba 2 space group symmetry. XRD pattern taken on mechanically cleaved crystal flake of synthesized crystal is shown in right inset of Fig. 2. High intense and sharp peaks are observed for some specific planes i.e. (002n) planes, where n=1,2,3,4. This shows that the synthesized crystal is grown uni-directionally along c-axis, confirming the single crystalline nature of synthesized crystal. Unit cell of synthesized $Sn_4Au$ drawn in VESTA software, is shown in left inset of Fig. 2. Each unit cell consists of two blocks of $Sn_4Au$ crystal, which is in accordance with the observed peaks for (002n) planes in XRD pattern taken on crystal flake.

Surface morphology of synthesized crystal is viewed through FESEM image, and is shown in Fig. 3(a). The synthesized crystal has typical terrace type morphology and the planes are arranged in a regular array. This shows the layer-by-layer growth of synthesized $Sn_4Au$ single crystal, confirming the crystalline nature of the same. EDS spectra is recorded to determine stoichiometry and purity of the sample and the same is shown in Fig. 3(b). EDS spectra shows that only two elements viz. Sn and Au are present in the synthesized sample and no other impurity element is seen. This shows the purity of the sample. Both the elements viz. Sn and Au are found to be in near stoichiometry ratio as shown in the inset of Fig. 3(b). Further, the single crystalline nature of the synthesized sample is verified by taking SAED pattern, as shown in Fig. 3(c). Typically, SAED pattern of a single crystal contains symmetrical spots due to diffraction from regularly arranged atomic planes. The SAED pattern of $Sn_4Au$ shows symmetrical diffraction spots, showing crystalline nature of the same. Spots from (002), (004), (006) and (008) planes are marked in Fig. 3(c). Fig. 3(d) shows the TEM image of $Sn_4Au$ single crystal, clearly exhibiting stacking of (002) atomic planes with a regular interval. These planes are separated by 2.95Å, which matches with the inter-atomic distance obtained for (002) planes in XRD measurement. The stacking of (002) planes as observed in TEM image is in agreement with XRD pattern taken on crystal flake, which also shows the unidirectional growth of crystal along (002n) direction.

XPS spectroscopy is used to determine chemical states of constituent elements of synthesized Sn4Au crystal. XPS peaks are calibrated with C 1s peak, which is considered as reference. XPS spectra recorded in Sn3d and Au4f regions is shown in Fig. 4(a) and 4(b). Gaussian fitting formula is used to fit the XPS peaks. Fig. 4(a) shows the fitted spin orbit doublet of Sn 3d core level. These peaks are deconvoluted in four peaks two for $SnO_2$ which corresponds to $Sn^{4+}$ state and the other two for $Sn^{2+}$ state [28,29]. Peaks for $SnO_2$ are observed due to air exposure of the sample. Peaks for $Sn^{2+}$ spin orbit doublets viz. Sn $3d_{5/2}$ and Sn $3d_{3/2}$ are found to be at 484.61±0.009eV and 493.05±0.008eV. The separation between these two peaks is found to be 8.44eV which is comparable with standard value 8.4eV [29]. Fig. 3(b) shows XPS spectra recorded in Au 4f region. This shows the peaks due to spin orbit doublet of Au 4f core levels viz. Au $4f_{7/2}$ and Au $4f_{5/2}$. These peaks are occurred at 84.91±0.004eV (Au $4f_{7/2}$) and at 88.59±0.005eV (Au $4f_{5/2}$). The separation between these peaks is found to be 3.68eV, which is comparable to standard value 3.67eV [29]. In ref. 29, XPS peak corresponding to Au $4f_{7/2}$ is specifically mentioned for Sn4Au compound. It is shown to be occurring in proximity of 85eV, and is observed in our case also. The observed values of XPS peaks in Au 4f regions confirm the formation of Sn4Au phase. The peak position along with the full width half maximum (FWHM) for various spin-orbit doublets of Sn and Au are given in table 2.

Fig. 5(a) depicts ρ-T measurements plot without field. A clear superconducting transition is observed with $T_c^{onset}$ at 2.6K, which is close to previously reported value of 2.5K [20,23,24]. Superconducting transition is sharp as can be seen in inset of Fig. 5(a). Zero resistivity is attained at 2.36K, providing a transition width of 0.25K. This signifies the quality of studied crystal. The ρ-T data is fitted in whole range above $T_c$ (5K-250K) using Bloch-Grüneisen formula. ρ(T), which is described by the following formalism,

$$\rho(T) = \left[\frac{1}{\rho_s} + \frac{1}{\rho_i(T)}\right]^{-1} \quad (1)$$

Here, $\rho_s$ represents temperature independent saturation resistivity and $\rho_i(T)$ is given by following equation,

$$\rho_i(T) = \rho(0) + \rho_{el\text{-}ph}(T) \quad (2)$$

here ρ(0) represents residual resistivity arising due to impurity scattering, the second term $\rho_{e\text{-}ph}(T)$ represents temperature dependent term, which depend on electron-phonon scattering. Further, $\rho_{e\text{-}ph}(T)$ is given by the following formula

$$\rho_{el-ph} = \alpha_{el-ph} \left(\frac{T}{\theta_D}\right)^n \int_0^{\frac{\theta_D}{T}} \frac{x^n}{(1-e^{-x})*(e^x+1)} dx \qquad (3)$$

here $\alpha_{el-ph}$ is electron-phonon coupling parameter, $\theta_D$ represents Debye temperature and n is constant. ρ-T data is well fitted with the above equation for n=5, signifying dominant electron-phonon scattering. Residual resistivity ratio (RRR) is found to be 19.06, which shows high metallicity of the synthesized crystal. The obtained value of Debye temperature $\theta_D$ from above fitting formula is 127.50±0.77K.

Fig. 5(b) shows the ρ-T-H measurements plot of synthesized $Sn_4Au$ single crystal. A clear suppression in $T_c$ is observed with applied magnetic field. It has been observed that onset of $T_c$ remains above 2K upto 120Oe, while zero resistivity is not achieved above 100Oe. Variation in critical field $H_c(0)$ is plotted against the temperature in inset of Fig. 5(b). This plot is fitted with conventional quadratic equation as given below

$$H_c(T) = H_c(0) * \left(1 - \frac{T^2}{T_c^2}\right) \qquad (4)$$

here, $H_c(0)$ is critical field at absolute zero, which is found to be 248Oe. Another approach to determine $H_c(0)$ is using Werthamer-Helfand-Hohenberg (WHH) formula, which is as follows:

$$H_c(0) = -0.693 T_c \left(\frac{dH_c}{dT}\right)_{T_c} \qquad (5)$$

linearly fitted plot is shown by blue curve in inset of Fig. 5(b), which is further extrapolated to T=0. $H_c(0)$ is estimated by multiplying extrapolated value by 0.693. Thus obtained $H_c(0)$ is found to be 285Oe.

Hall resistivity ($\rho_{xy}$) is measured with respect to applied field in a range of ±4T at 3K and the same is shown in Fig. 5(c). Field dependent Hall resistivity gives the information about carrier density and nature of the carriers. $\rho_{xy}$ is found to have linear relationship with applied magnetic field. Hall coefficient is determined by taking the slop of linearly fitted $\rho_{xy}$ vs H plot. Hall coefficient is found to be -4.05±0.02×$10^{-10}$Ω-m-$T^{-1}$. The negative sign of Hall coefficient shows that the electrons are the dominant charge carriers. The charge carrier density is calculated by using the formula, $R_H = -\frac{1}{ne}$, where charge carrier density is given by n and e is the electronic charge. The calculated value of charge carrier density is found to be 1.54±0.03×$10^{28}m^{-3}$. Further, this value is verified by calculating charge carrier density theoretically. $Sn_4Au$ unit cell contains 4 formula units, in which each formula unit provides 2 electrons. In this way, $Sn_4Au$ has 8 electrons per unit cell. The total electron density is calculated by dividing number of electrons per unit cell by volume of unit cell. Volume of $Sn_4Au$ unit cell obtained from Rietveld refinement is V=498.783Å$^3$. The electron density is found to be $1.60 \times 10^{28}$m$^{-3}$, this value closely matches with the value obtained from the experiment. The value of charge carrier density is used to determine Fermi wave vector by considering a spherical Fermi surface. Fermi wave vector is given by $k_F=(3n\pi^2)^{1/3}$, where n is charge carrier density. Here, the value of n obtained from Hall calculations is used, and the corresponding Fermi wave vector $k_F$ is found to be 0.78Å$^{-1}$. These values are used to calculate the superconductivity characteristic lengths viz. London penetration depth $\lambda(0)$ and BCS coherence length $\xi(0)$. The value of London penetration depth $\lambda(0)$ can be calculated by using the relation $\lambda(0) = \left(\frac{m^*}{\mu_0 n e^2}\right)^{1/2}$, here m$^*$ is taken as free electron mass for simplicity. The obtained value of $\lambda(0)$ is 42.86nm. The BCS coherence length is calculated using the formula $\xi(0) = \frac{0.18\hbar^2 k_F}{k_B T_c m^*}$, and it is found to be 480.7nm. The values of both the superconducting lengths are used to determine Ginzberg Landau (GL) κ parameter. κ parameter is given by $\kappa = \frac{\lambda(0)}{\xi(0)}$, and its value is found to be 0.08, which is less than cut-off value for Type-I superconductivity. This shows $Sn_4Au$ to be a Type-I superconductor. κ parameter can also be determined by calculating the value of mean free path. The mean free path (l) can be calculated using the formula $l = v_F \tau$, here $v_F$ is Fermi velocity and given by $v_F = \frac{\hbar k_F}{m^*}$, and τ is scattering time and given by $\tau = \frac{m^*}{n e^2 \rho}$, here ρ is taken as residual resistivity ρ=1.08×10$^{-7}$Ω-m. The value of mean free path (l) is found to be 119.2nm. Certainly, $\xi(0)$ is very large as compared to the mean free path, suggesting $Sn_4Au$ to be a dirty limit superconductor. For a dirty limit superconductor, Ginzberg Landau (G-L) κ parameter can be calculated as κ=$\frac{0.75\lambda(0)}{l}$, the obtained value of κ is 0.27 which is lower than the limit of type-I superconductivity, suggesting $Sn_4Au$ to be a type-I superconductor. In ref. 20, $Sn_4Au$ is shown to have type-II superconductivity with relatively lower critical field ($H_{c1}$) and upper critical field ($H_{c2}$). In this article $Sn_4Au$ is seen to be a type-I superconductor based on κ parameter obtained from superconductivity characteristics lengths. Various superconducting parameters for present selfflux grown $Sn_4Au$ are given in table 3.

Fig. 6(a) shows ρ vs H plot of $Sn_4Au$ single crystal at 2K, 3K, 5K and 10K in a field range of ±0.8T. At 2K, sample is in its superconducting state and its resistivity remains zero up to 120Oe, and sharply increases when the field is further increased. This shows $H_c$ to be 120Oe at 2K, which is in accordance with the ρTH measurements. ρ-H plots show no indication of superconductivity at 3K, 5K and 10K. An interesting feature is observed in terms of increase of resistivity with field, hinting towards significant MR in normal state. The study of normal state of a topological superconductor is very crucial to determine normal state topological properties of the same. To analyse normal state properties of synthesized $Sn_4Au$ single crystal, ρ-H measurements are carried out in a field range of ±12T in normal state i.e., at temperatures 3K, 5K, 10K and 20K, which are shown in Fig. 6(b). MR % has been calculated by using the following formula

$$MR\% = \frac{\rho(H)-\rho(0)}{\rho(0)} \times 100 \qquad (6)$$

MR% is found to be around 150%, 120%, 110% and 65% at 3K, 5K, 10K and 20K respectively. This positive MR in normal state, indicates towards possible surface states driven transport in $Sn_4Au$ single crystal. MR seems to be linear and non-saturating in complete magnetic field range. There can be two possible reasons for the observed linear magnetoresistance (LMR). The first one has the classical origin and emerges due to impurity scattering [30] in an inhomogeneous sample. This reason is not valid here as the synthesized $Sn_4Au$ crystal is highly crystalline and there is no indication of inhomogeneity in the sample. The other reason for LMR has the quantum origin, which is observed in topological materials [31-33]. In topological materials, occurrence of LMR is attributed to the surface states dominated transport phenomenon, while a deviation from linear to quadratic dependency of MR on magnetic field indicates bulk dominated transport properties. Here, high field MR at 3K is fitted with power law i.e. MR%=$kH^\gamma$ (inset of Fig. 6(b)), where the value of exponent γ determine whether the conduction is dominated by bulk states or the topological surface states. The value of γ near to 1 (linear) shows surface states dominated conduction mechanism while the value of γ near to 2 (quadratic) shows that the conduction is dominated by bulk states and has the classical origin. For synthesized $Sn_4Au$ single crystal, this value of γ is found to be 0.8, which is close to 1, and shows that the conduction is dominated by topological surface states [33]. Interestingly a V type cusp has also been observed in low field MR data, which signifies possible existence of WAL effect in synthesized $Sn_4Au$ single crystal [34]. This V type cusp broadens at higher temperatures, this happens due to shortening of coherence length $l_\phi$. This possible WAL effect is verified by analysing magnetoconductivity at different temperatures.

In Fig. 6(c), magnetoconductivity (MC) is plotted against the applied field in a temperature range of ±1T at 3K, 5K, 10K and 20K. Presence of WAL effect in a topological material, can be confirmed by fitting low temperature MC with HLN formalism [35]. In HLN formalism, the difference in magnetoconductivity $\Delta\sigma(H)$ is taken as $[\sigma(H)-\sigma(0)]$, and the same can be described with the following equation

$$\Delta\sigma(H) = -\frac{\alpha e^2}{\pi h}\left[\ln\left(\frac{B_\varphi}{H}\right) - \Psi\left(\frac{1}{2} + \frac{B_\varphi}{H}\right)\right] \tag{7}$$

Here, $B_\phi$ is the characteristic field and is given by $B_\varphi = \frac{h}{8e\pi l_\varphi^2}$, $l_\phi$ is phase coherence length and $\Psi$ is digamma function. The phase coherence length $l_\phi$ is defined as the maximum length travelled by electron while maintaining its phase. The value of pre-factor $\alpha$ determines which localization is present in the system, a positive value of $\alpha$ signifies weak localization (WL) effect, while a negative value of $\alpha$ indicates the presence of WAL effect. In topological materials, WAL is supposed to be induced by topological surface states. These topological surface states are protected by $\pi$ Berry phase, and the presence of $\pi$ Berry phase. For a material with $\pi$ Berry phase, the value of pre-factor $\alpha$ should be -0.5 [34]. Intriguingly, the value of $\alpha$ depends upon number of topological surface states (TSS), for each TSS, $\alpha$ takes the value to be -0.5. If $\alpha$ takes the value to be equal to -1, it shows that there are two distinct TSS are present in the system, one is at the top and other at the bottom [34,36]. The value of $\alpha$ between -0.5 and -1, shows that the top and bottom surface states are connected through bulk conducting channels [37]. This suggest that the conduction is contributed by the both viz. bulk conducting channels and topological surface states. In Fig. 6(c), the solid black curves show the HLN fitting results at 3K, 5K, 10K and 20K. The obtained value of pre-factor $\alpha$ and phase coherence length $l_\phi$ are listed in table-4. The value of $\alpha$ at 3K is found to be -1.028, which is near to -1, and suggest that two distinct TSS contributes to the conduction mechanism in synthesized $Sn_4Au$ single crystal. Also, it has been observed that as the temperature is increased, the value of $\alpha$ is decreased, and it is found to be -0.2966 at 20K as shown by encircled dots in Fig. 6(d). The deviation of the value of $\alpha$ from the standard values viz. -0.5 and -1, suggest that bulk states start to contribute in conduction at higher temperatures. The variation of inverse of square of phase coherence length $l_\phi^{-2}$ with respect to temperature is shown by blue symbols in Fig.

6(d). Temperature dependence of $l_\phi$, gives the information about the scattering process and dephasing mechanism. According to Nyquist theory, if $l_\phi^{-2}$ varies linearly with temperature, this shows that there is only electron-electron (e-e) scattering present in the system [38,39]. At low temperatures, e-e scattering is the only dephasing mechanism, while at higher temperatures this scattering is assisted by electron-phonon (e-p) scattering [39]. In Fig. 6(d), the solid red line shows linearly fitted $l_\phi^{-2}$ vs T plot, and it is clear that $l_\phi^{-2}$ is not well fitted with linear equation, suggesting the presence of both scattering processes viz. e-e and e-p. To determine nature of WAL effect, whether it is 2D or 3D, the $l_\phi^{-2}$-T plot is fitted with the power law as follows [40],

$$\frac{1}{l_\phi^2(T)} = \frac{1}{l_\phi^2(0)} + A_{e-e}T^p + A_{e-p}T^q \tag{8}$$

Here, $l_\phi(0)$ is the dephasing length at absolute zero and $A_{e-e}T^p$ and $A_{e-p}T^q$ represents the contribution from e-e scattering and e-p scattering respectively. The fitted plot is shown with solid black line in Fig. 6(d). Here, the value of p and q are found to be 1 and 2 respectively. The values of $A_{e-e}$ and $A_{e-p}$ are found to be $3.101 \times 10^{-6}$ and $3.30 \times 10^{-7}$. This result shows that both the scattering process viz. e-e and e-p scattering play a role in dephasing mechanism of $Sn_4Au$. Also, the obtained values of p and q show that the both e-e and e-p scattering are 2D in nature. This signifies the presence of 2D WAL effect in synthesized $Sn_4Au$ single crystal. This because in 3D WAL effect, 3D e-p scattering dominates, and the power law is changed from $l_\phi^{-2} \propto T^2$ to $l_\phi^{-2} \propto T^3$ [39,41]. The WAL effect can be originated due to two reasons, the first one is the existence of TSS in case of 2D and the second one is the high spin orbit coupling (SOC) in the bulk states of the system in case of 3D. To confirm TSS to be the reason of observed 2D WAL effect, angle dependent magneto transport measurements have been carried out, in which angle between the applied field and direction of current is varied.

Fig. 7(a) is showing anisotropic magnetoresistance (AMR) of synthesized $Sn_4Au$ single under a magnetic field of 12T at 3K. The transverse field direction is taken as the initial position and tilt angle is measured by taking the same as reference or 0° position. The geometry of current and field direction is shown in inset of Fig. 7(b). Resistivity at 12T, is found to be periodic with the tilt angle, with regular occurring maxima and minima. Maxima of resistivity is obtained when the field is perpendicular to current (at 0° and 180° tilt angle) and minima is obtained when the field is parallel to current (at 90° and 270° tilt angle). It has been well explained in literature that the TSS dominated electric transport depends only on the perpendicular component of the field [42,43]. This periodicity of MR, with maxima values at perpendicular field shows that the electrical transport is governed by TSS. Further, MR% has been calculated at different tilt angles viz. 0°, 30°, 45°, 60°, 75° and the same is shown in Fig. 7(b). MR% is found to decrease as the tilt angle is increased. Also, the sharp V-type cusp at lower magnetic field starts to flatten with increment in tilt angle. This shows that WAL is suppressed with increased tilt angle. Angle dependent magnetoconductivity is supposed to be the most prominent source to determine whether the observed WAL effect is due to TSS or the strong SOC of bulk of material [44-46]. To ascertain the same, in Fig. 7(c), the normalized magnetoconductivity is plotted against the applied magnetic field at 3K with various tilt angles viz. 0°, 30°, 45°, 60° and 75°. Further, the normalized magnetoconductivity is plotted against Hcosθ, and the same is shown in Fig. 7(d). All normalized MC plots are found to merge in a single plot at low magnetic fields, this confirms that the observed WAL is 2D in nature and arises due to the presence of TSS in synthesized $Sn_4Au$ single crystal [37,44]. For a material, in which WAL effect arise due to strong bulk spin orbit coupling, normalized MC vs Hcosθ plots are found to be well separated from each other [45,46]. Our angle dependent magneto-transport measurements result support the findings of HLN model, showing that the electrical transport in synthesized $Sn_4Au$ single crystal is governed by topological surface states at low temperature and low magnetic fields. More recently, in theoretical calculations, $Sn_4Au$ is shown to have topological non-trivial bulk electronic band structure and the presence of topology is also verified by calculating $Z_2$ invariants [26]. Our report along with the previous report on $Sn_4Au$ [20], showed the presence of TSS through magneto transport measurements. This is yet to be verified by direct measurements such as angle resolved photo electron spectroscopy (ARPES). The, ARPES measurements on good quality single crystals of $Sn_4Au$, to get more confidence on presence of topological surface states in $Sn_4Au$ are thus warrented.

**Conclusion:**

In conclusion, single crystal of $Sn_4Au$ is synthesized through a relatively easy flux free method following the solid state reaction route and its quality is verified through host of characterization techniques viz. XRD, FESEM, TEM and XPS. Superconducting properties of $Sn_4Au$ being studied through ρ-T and ρTH measurements, showed the same to be a type-I superconductor. Detailed ρTH measurements being examined though detailed HLN analysis of MC resulted in occurrence of WAL effect due to TSS dominated transport phenomenon. This is further, verified through normalized MC vs Hcosθ at different tilt angle. This report shows that $Sn_4Au$ shows superconducting properties below $T_c$ as well TSS dominated topological properties in its normal state. This unique feature of Sn$_4$Au establish this material as a good choice to get more understanding in the field of topological superconductivity.

**Acknowledgment:**

Authors would like to thank Director of National Physical Laboratory, New Delhi for his keen interest. Authors would like to thank Dr. R.N. Bhowmik, Pondicherry University and CIF of Pondicherry University for XPS measurements. M. M. Sharma would like to thank CSIR, India for research fellowship and AcSIR, Ghaziabad for Ph.D. registration.

**Table-1**

Parameters obtained from Rietveld refinement:

| Cell Parameters | Refinement Parameters |
|---|---|
| Cell type: Face Centred Cubic (FCC) | $\chi^2$=1.82 |
| Space Group: F m -3 m | $R_p$=6.14 |
| Lattice parameters: a=6.514(5)Å | $R_{wp}$=7.93 |
| b=6.529(1)Å & c=11.726 Å | $R_{exp}$=5.87 |
| $\alpha=\beta=\gamma=90°$ | |
| Cell volume: 498.783Å$^3$ | |
| Density: 8.945 g/cm$^3$ | |
| Atomic co-ordinates: | |
| Sn1 (0.1694,0.3395,0.1242) | |
| Au (0,0,0) | |
| Sn2 (0.3502,0.1624,0.8574) | |

**Table-2**

XPS peaks position and FWHM of constituent elements of synthesized Sn$_4$Au single crystal:

| Element | Spin-orbit doublet | Binding Energy | FWHM |
|---|---|---|---|
| Sn | 3d$_{5/2}$ | 484.61±0.009eV | 0.76±0.03eV |
| | 3d$_{3/2}$ | 493.05±0.008eV | 0.77±0.04eV |
| Au | 3d$_{5/2}$ | 84.91±0.004eV | 0.82±0.09eV |
| | 3d$_{3/2}$ | 88.59±0.005eV | 0.81±0.01eV |

### Table-3

Parameters obtained from Heat capacity measurements:

| Parameter | Obtained Value |
|---|---|
| Superconducting critical temperature ($T_c^{onset}$) | 2.6K |
| Debye temperature ($\theta_D$) | 127.5±0.77K |
| Critical field at absolute zero, $H_c(0)$ | 248Oe |
| Charge carrier density ($n_e$) | 1.54±0.03×$10^{28}m^{-3}$ |
| London penetration depth $\lambda(0)$ | 42.86nm |
| BCS coherence length $\xi(0)$ | 480.7nm |
| Mean free path (l) | 119.2nm |

### Table: 4

Low field (up to 1 Tesla) HLN fitted parameters of $Sn_4Au$ single crystal

| Temperature(K) | α | $l_\phi$ |
|---|---|---|
| 3 | -1.028 | 113.0303 |
| 5 | -0.7973 | 101.6673 |
| 10 | -0.7473 | 87.4037 |
| 20 | -0.2966 | 61.6531 |

**Figure Captions:**

**Fig. 1:** Schematic of heat treatment followed to synthesize single crystal of $Sn_4Au$ through simple solid state reaction route.

**Fig. 2:** Rietveld refined PXRD pattern of synthesized $Sn_4Au$ single crystal in which, right inset is showing the XRD pattern taken on crystal flake and the left inset is showing the VESTA drawn unit cell of $Sn_4Au$.

**Fig. 3(a):** SEM image of synthesized $Sn_4Au$ single crystal (b) EDAX spectra of synthesized $Sn_4Au$ single crystal in which inset is showing the atomic composition of the same (c) SAED pattern of synthesized $Sn_4Au$ single crystal (d) TEM image of synthesized $Sn_4Au$ single crystal.

**Fig. 4:** XPS spectra in (a) Sn 3d region (b) Au 4f region.

**Fig. 5(a):** B-G fitted $\rho$-T plot of synthesized $Sn_4Au$ single crystal in which inset is showing the zoomed view of $\rho$-T plot in the proximity of transition temperature. (b) $\rho$–T plots of synthesized $Sn_4Au$ single crystal at different applied fields, in which inset is showing the fitted plot of $H_c$ vs T. (c) Hall resistivity $\rho_{xy}$ vs H plot of synthesized $Sn_4Au$ single crystal at 5K.

**Fig. 6(a):** Resistivity vs Field plot at 2K, 3K, 5K and 10K. (b) MR% vs H plot at 3K, 5K, 10K and 20K od synthesized $Sn_4Au$ single crystal, in which inset is showing the power law fitted high field MR% at 3K. (c) Low field (±1T) HLN fitted MC of synthesized $Sn_4Au$ single crystal. (d) Variation of parameters obtained from HLN fitting with respect to temperature, wine-coloured symbols are showing variation of $\alpha$ with respect to temperature, while blue symbols are showing variation of $l_\phi^{-2}$ with respect to temperature. Solid red and black curve are showing linear fitted and power law fitted $l_\phi^{-2}$ vs T plot respectively.

**Fig. 7(a):** Variation is resistivity with respect to the tilt angle in a field of 12T and at 3K of synthesized $Sn_4Au$ single crystal. (b) Magnetoresistance vs H plot of synthesized $Sn_4Au$ single crystal at 3K in a field range of ±12T at various tilt angles, in which inset is showing the geometry used for the measurements (c) Normalized MC vs H plot of synthesized $Sn_4Au$ single crystal at 3K in a field range of ±12T at various tilt angles. (d) Normalized MC vs $H\cos\theta$ plot of synthesized $Sn_4Au$ single crystal at 3K in a field range of ±12T.

Fig. 1

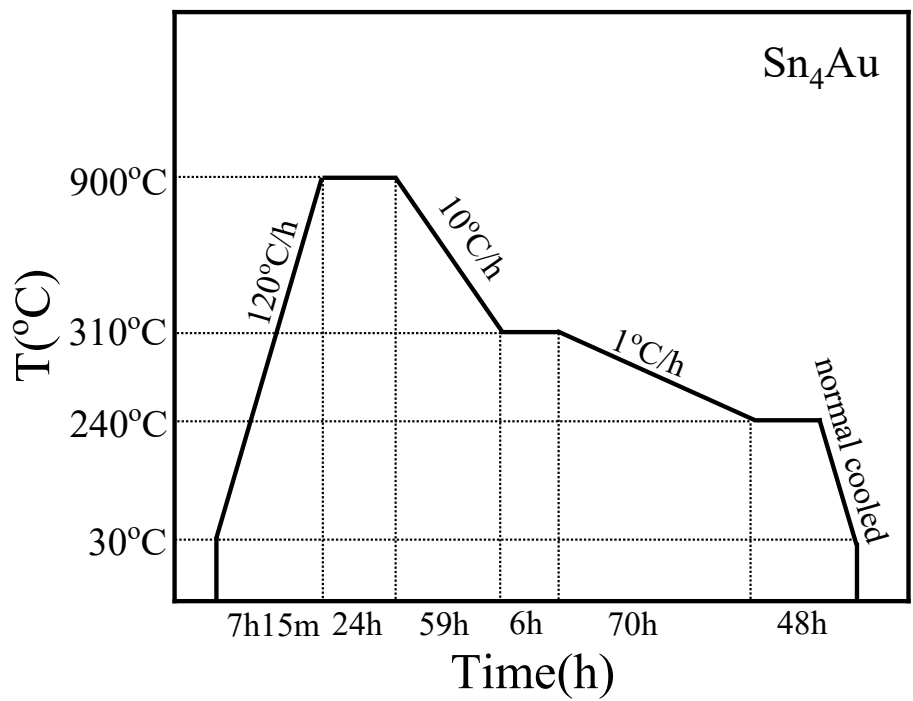

Fig. 2

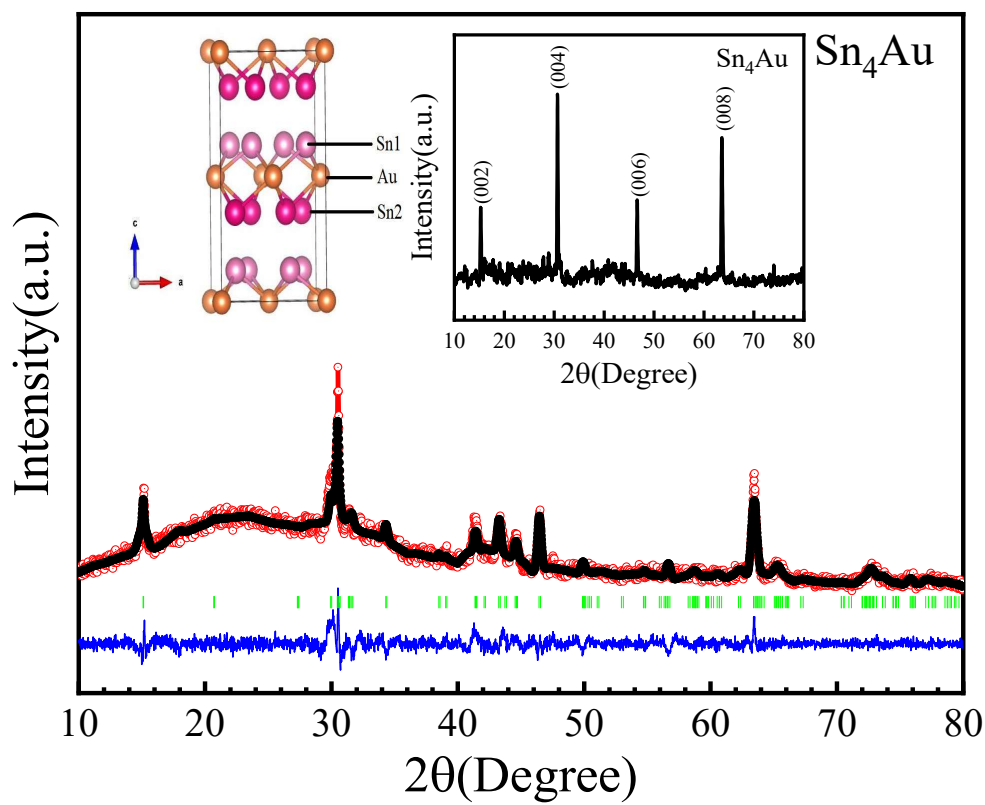

Fig. 3

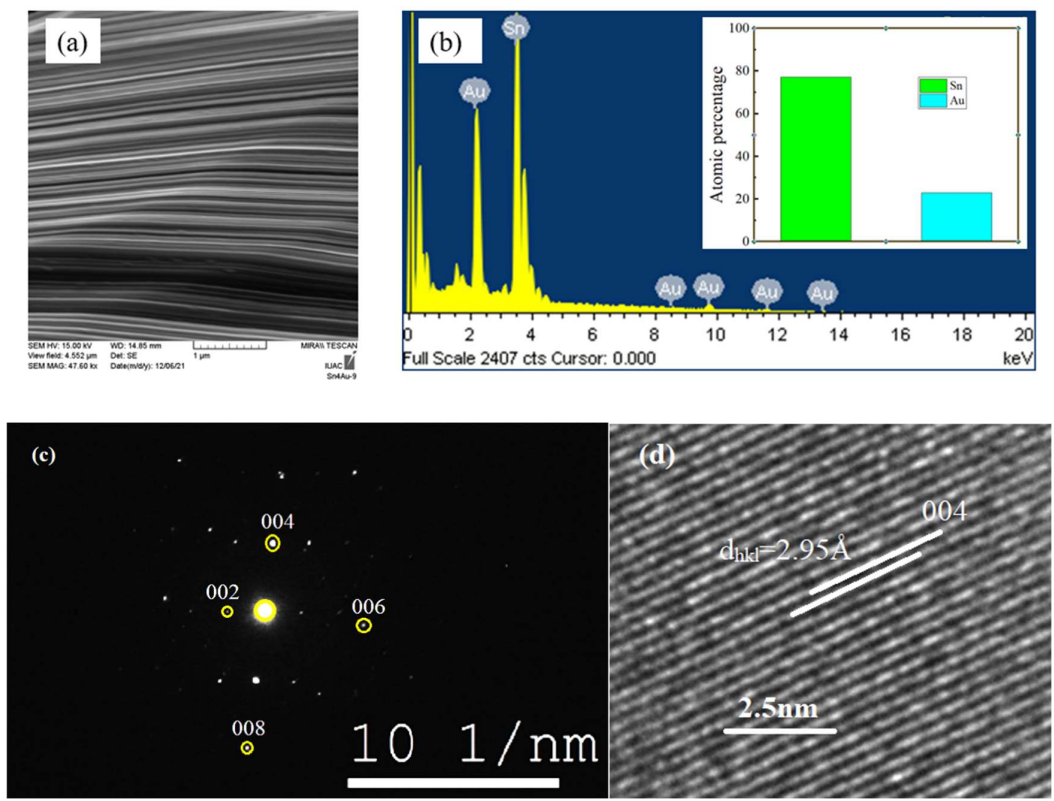

Fig. 4(a)&(b)

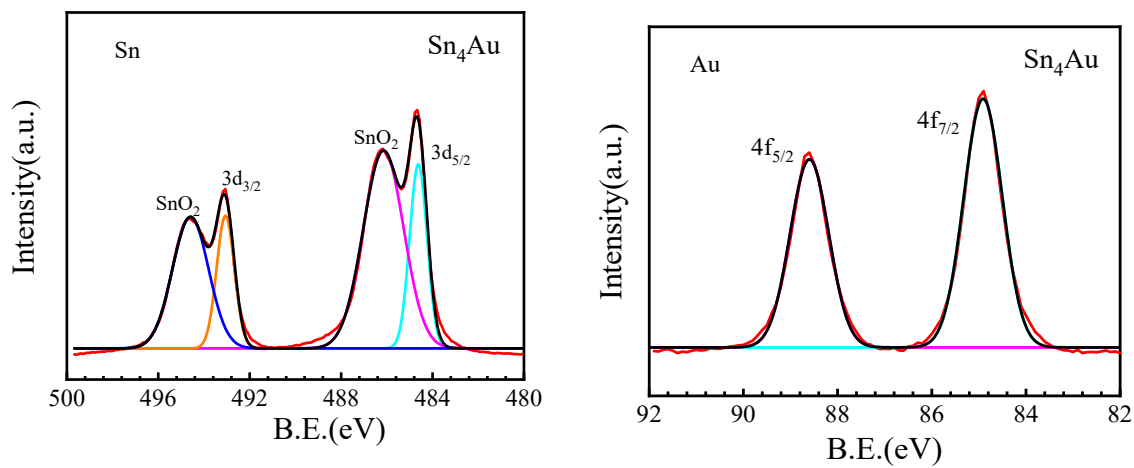
19

Fig. 5(a)

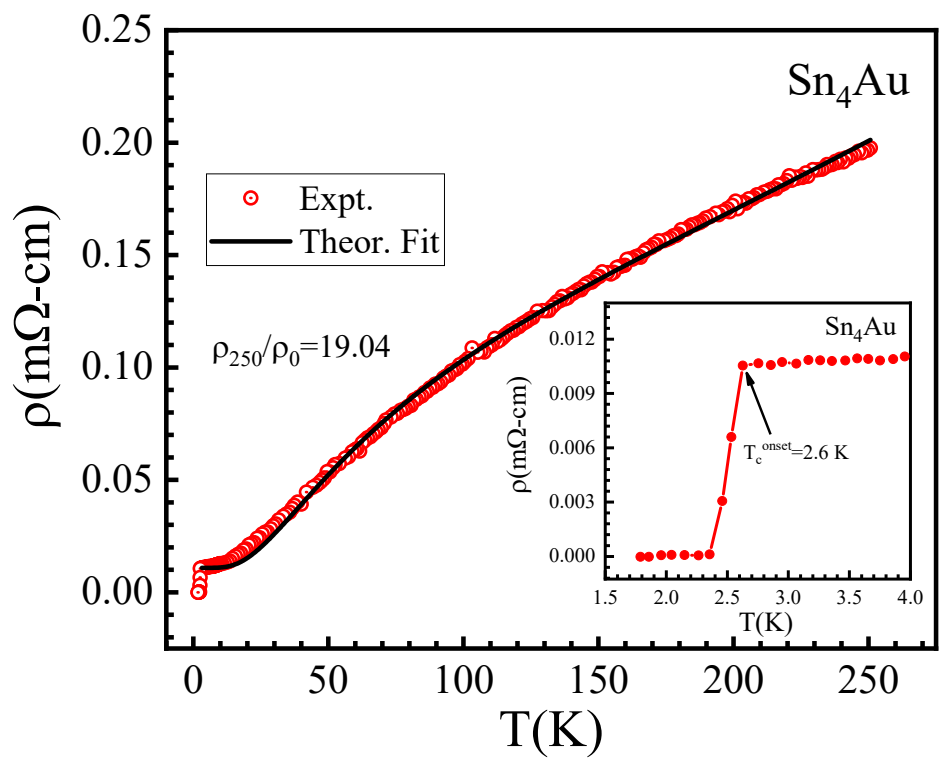

Fig. 5(b)

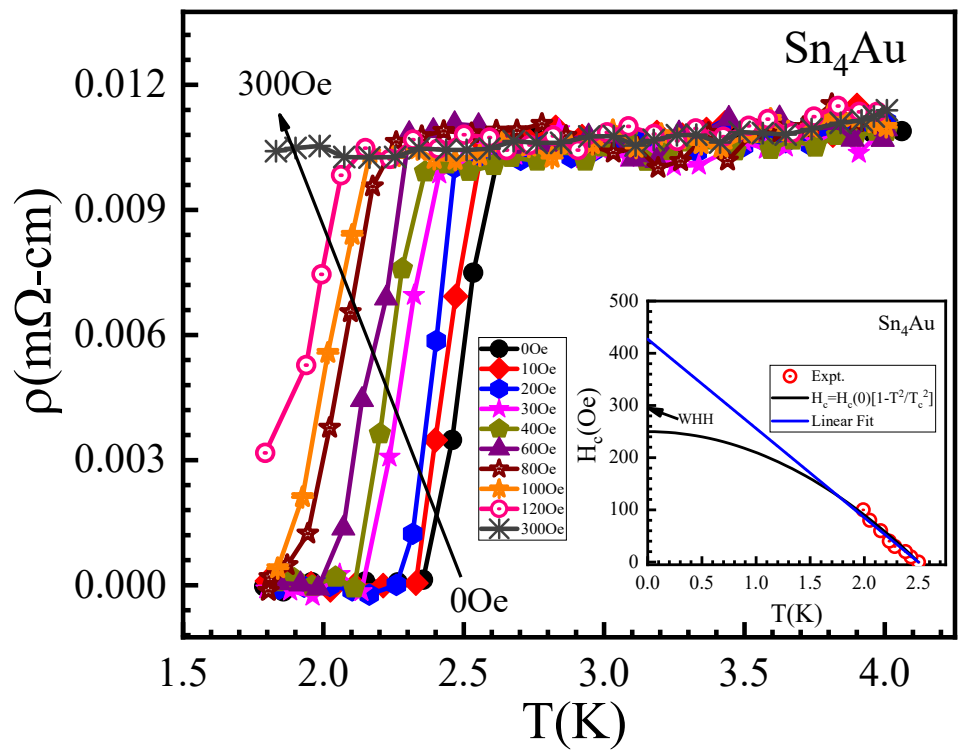

Fig. 5(c)

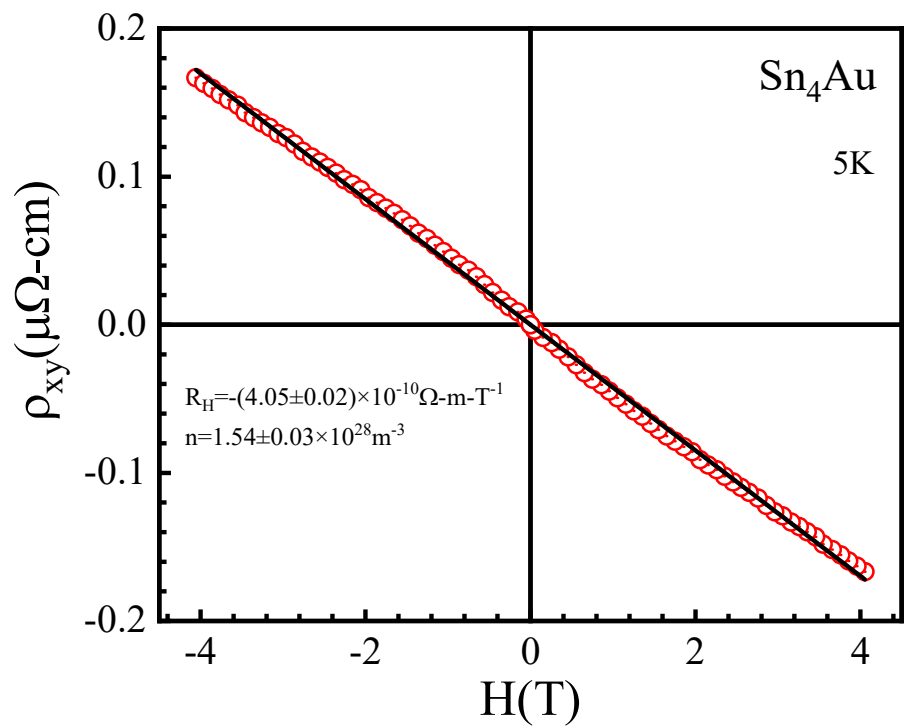

Fig. 6(a)

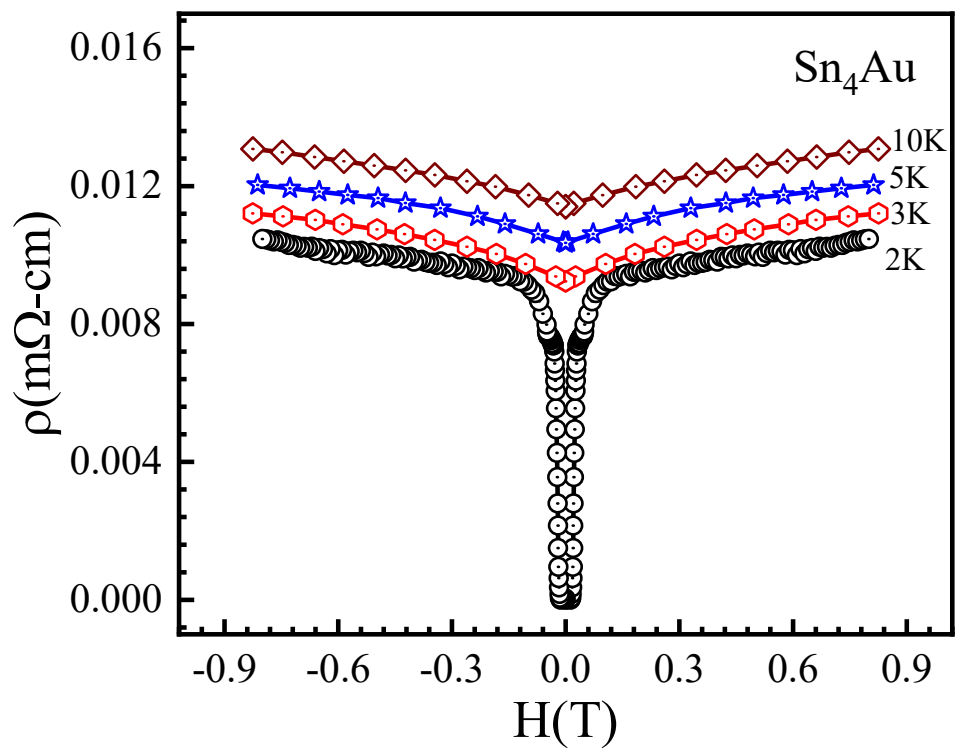

Fig. 6(b)

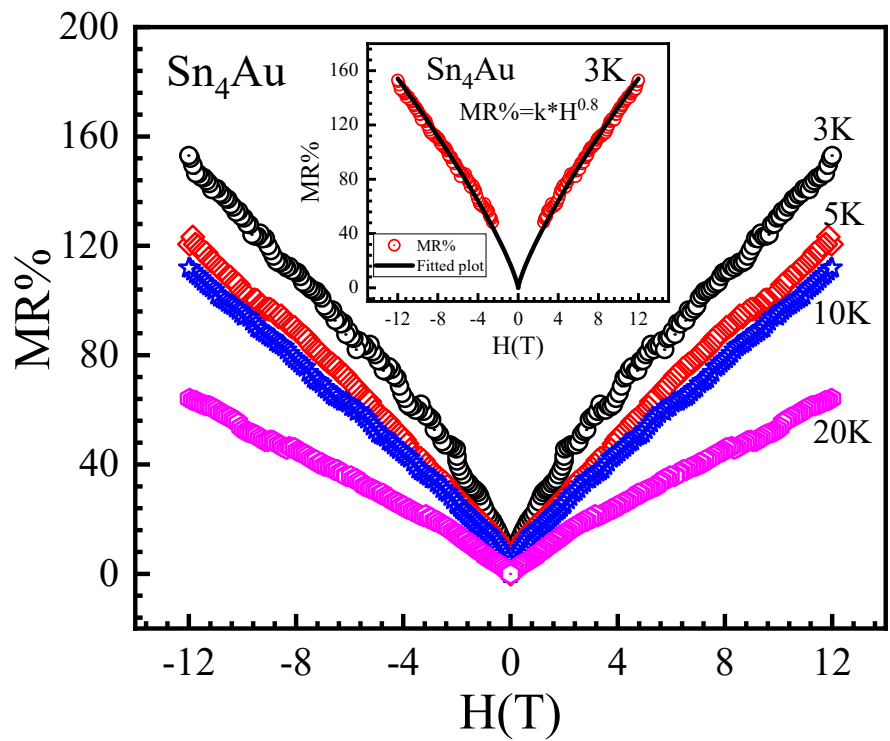

Fig. 6(c)

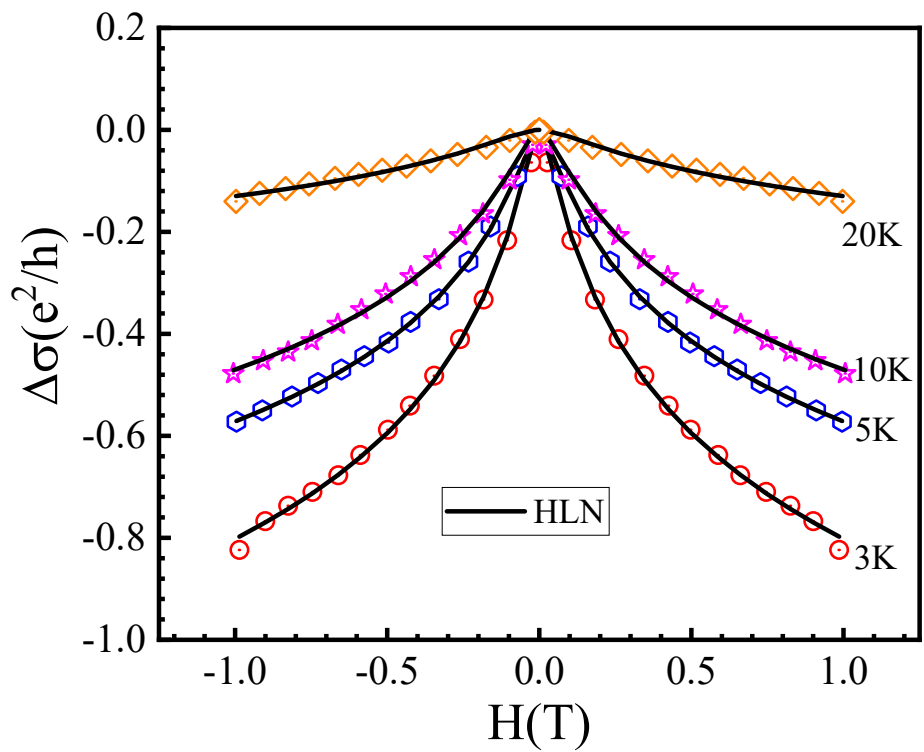

Fig. 6(d)

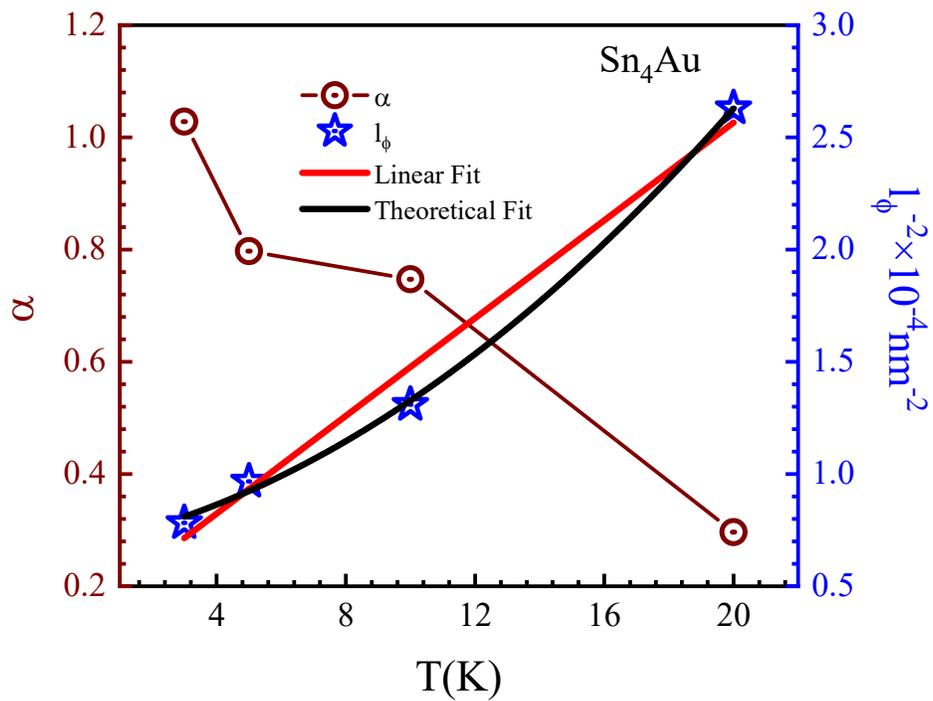

Fig. 7(a)

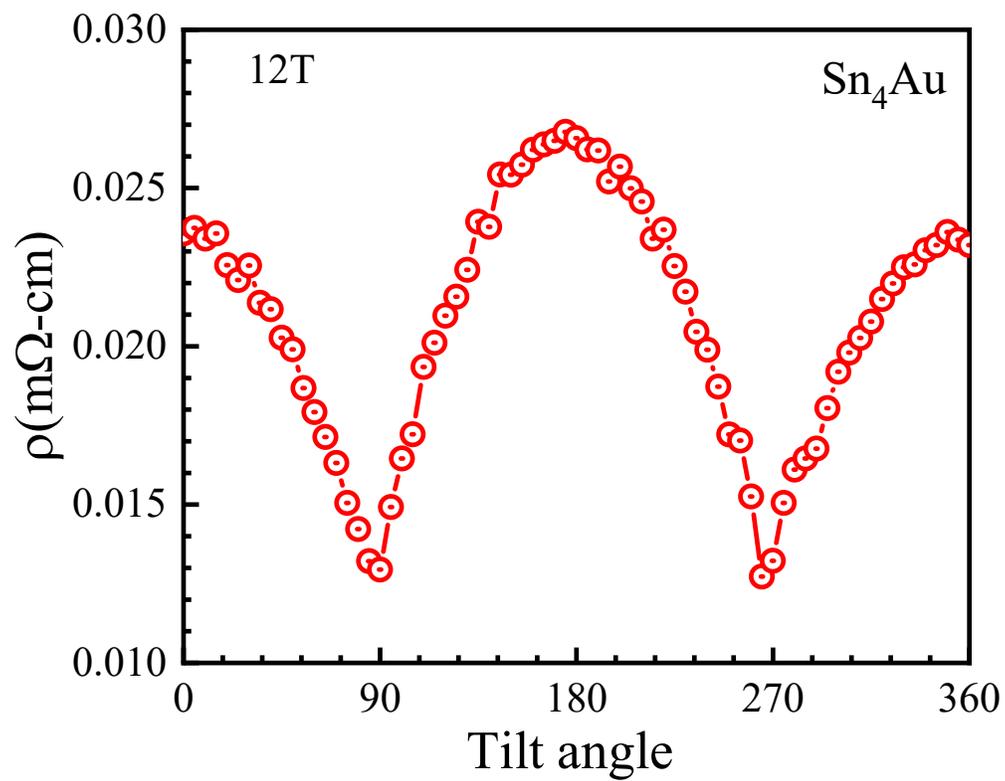

Fig. 7(b)

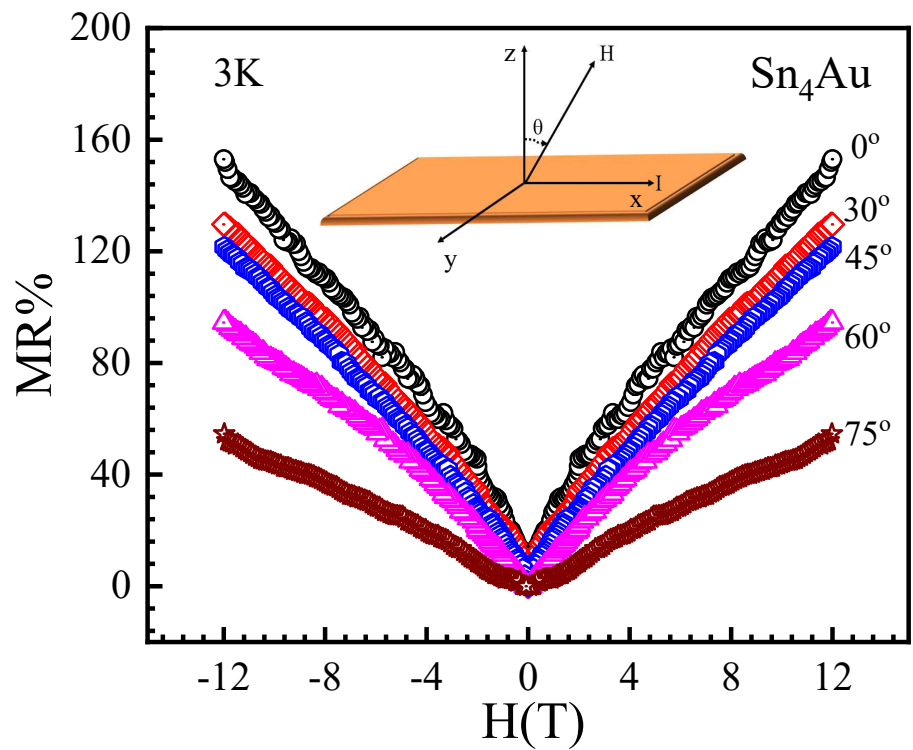

Fig. 7(c)

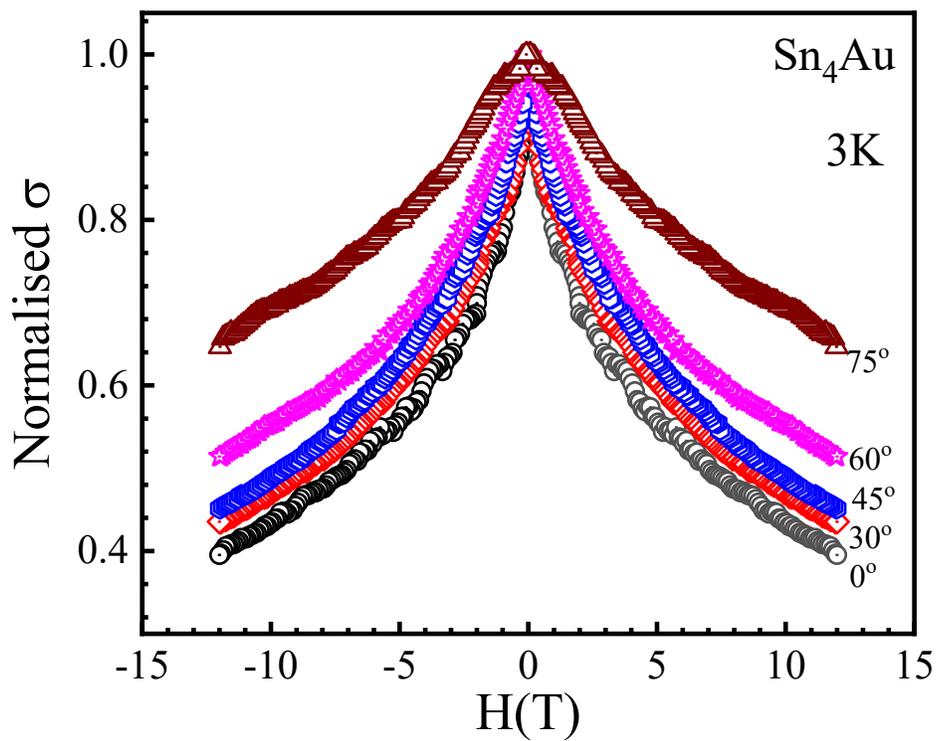

Fig. 7(d)

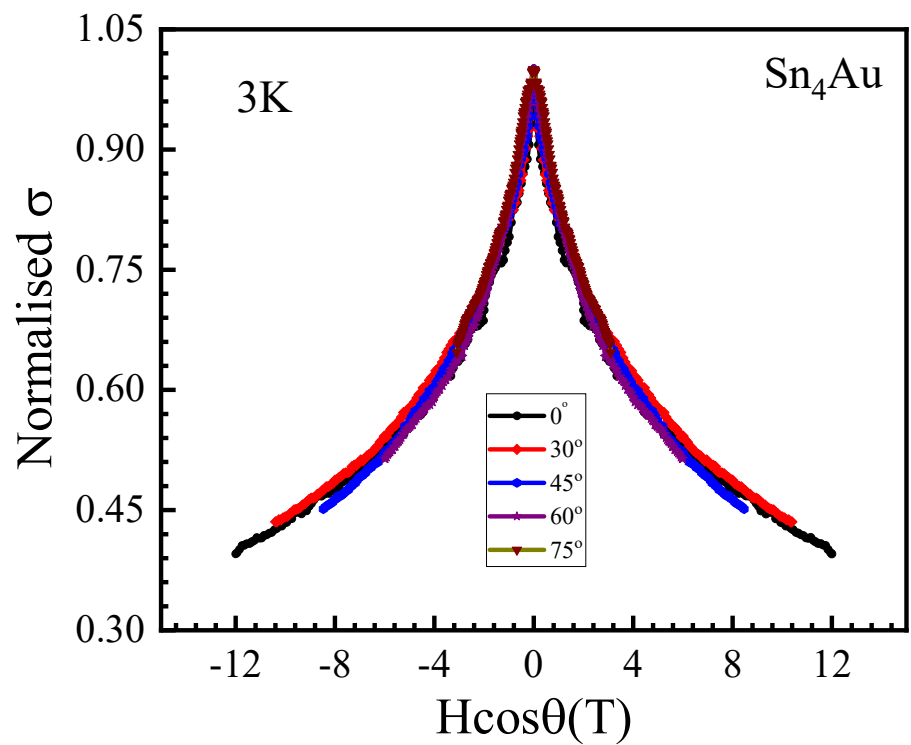